\documentclass[aip,pop,amsmath,amssymb,reprint]{revtex4-1}

\usepackage{graphicx}
\usepackage{dcolumn}
\usepackage{bm}

\usepackage[utf8]{inputenc}
\usepackage[T1]{fontenc}
\usepackage{newtxtext}
\usepackage[varg]{newtxmath} 
\usepackage{etoolbox}

\usepackage[english]{babel}
\usepackage[autostyle]{csquotes}
\usepackage{parskip}
\usepackage{amsmath}
\usepackage{derivative}
\usepackage{amsfonts}
\usepackage{amssymb}
\usepackage{titlesec}
\usepackage{indentfirst}
\usepackage{multirow}
\usepackage{hyperref}
\usepackage{xcolor}
\newcommand{\hide}[1]{}

\newcommand{\bB}{{\bf B}}
\newcommand{\bD}{{\bf D}}

\newcommand{\bM}{{\bf M}}
\newcommand{\bR}{{\bf R}}
\newcommand{\ba}{{\bf a}}

\newcommand{\bu}{{\bf u}}
\newcommand{\bv}{{\bf v}}

\newcommand{\cC}{{\cal C}}
\newcommand{\cJ}{{\cal J}}

\newcommand{\cO}{{\cal O}}

\draft 

\begin{document}


\title{Axisymmetric Gyrokinetic Simulation of ASDEX-Upgrade Scrape-off Layer Using a Conservative Implicit BGK Collision Operator} 



\author{D. Liu}
\email[]{dingyunl@princeton.edu}
\affiliation{Princeton University}

\author{J. Juno}
\affiliation{Princeton Plasma Physics Laboratory}

\author{G. W. Hammett}
\affiliation{Princeton Plasma Physics Laboratory}

\author{A. Hakim}
\affiliation{Princeton Plasma Physics Laboratory}

\author{A. Shukla}
\affiliation{University of Texas at Austin}

\author{M. Francisquez}
\affiliation{Princeton Plasma Physics Laboratory}


\date{\today}

\begin{abstract}
Collisions play an important role in turbulence and transport of fusion plasmas. For kinetic simulations, as the collisionality increases in the domain of interest, the size of the time step to resolve the collisional physics can become overly restrictive in an explicit time integration scheme, leading to high computational cost. With the aim of overcoming such restriction, we have implemented an implicit Bhatnagar–Gross–Krook (BGK) collision operator for use in the discontinuous Galerkin (DG) full-\textit{f} gyrokinetic solver within the Gkeyll framework, which, when combined with Gkeyll's traditional explicit time integrator for collisionless advection, can significantly increase the time step in gyrokinetic simulations of highly collisional regimes. To ensure conservation of density, momentum, and energy, we utilize an iterative scheme to correct the discretized approximation to the equilibrium Maxwellian distribution to which the BGK collision operator relaxes. We have further generalized the BGK infrastructure, both the implicit scheme and the correction routine, to handle cross species collisions. This improved implicit and conservative BGK operator is benchmarked against the more accurate but more computationally expensive Lenard-Bernstein-Dougherty (LBD) operator which has been utilized in prior studies with Gkeyll. The implicit BGK operator enables 2D axisymmetric simulations of the ASDEX-Upgrade scrape-off layer to run $56$ times faster to completion than the simulations with the LBD operator, because the BGK operator is more robust and converges at a lower resolution than is required by the LBD operator. Additionally, in this more collisional limit, we demonstrate that the results of our simulations utilizing the implicit BGK operator agreed well with simulations utilizing the more computationally expensive LBD operator.
\end{abstract}


\maketitle 


\section{\label{sec: intro}Introduction}
Understanding heat and particle transport in the scrape-off layer (SOL) of tokamaks is important to better handle power deposition onto divertor plates. Fluid codes such as SOLPS-ITER \cite{Schneider_1992}, UEDGE \cite{Rognlien_1992}, SOLEDGE2D-EIRENE \cite{Bufferand_2015}, and EDGE2D-EIRENE \cite{Radford_1996} are commonly used to simulate 2D transport in the SOL. The fluid equations are a good approximation to describe plasma dynamics in the highly collisional limit, where the electron and ion distributions are close to thermodynamic equilibrium, i.e., a Maxwellian distribution. However, while the SOL collisionality is high compared to that in the core, the collisionality is not always sufficiently high for kinetic effects to become negligible. Even when the collision frequency in the SOL or near the divertor plates is high enough that fluid equations are rigorous, the collision frequency in the pedestal and edge region inside the last-closed flux surface can be low enough that kinetic effects are important, so we need a code that can handle both kinetic and collisional regimes simultaneously. 
In the nonuniform magnetic field of a tokamak, particles are subject to drifts, and mirror trapping can alter parallel transport in the presence of ion temperature anisotropy. It has often been numerically hard to treat drifts accurately in fluid codes, while a kinetic solver naturally includes the drifts. Therefore, we are interested in including those kinetic effects by running gyrokinetic simulations. 

For this study, we will utilize the full-\textit{f} gyrokinetic solver in the Gkeyll framework \cite{Francisquez_2025, HakimJuno_2020}. By virtue of being full-\textit{f}, Gkeyll is capable of dealing with large-amplitude perturbations in the SOL. Here, we focus on the long wavelength limit of the gyrokinetic equation, which has previously been applied to turbulence studies in linear \cite{Shi_thesis}, helical \cite{Shi_2019, Bernard_2020}, and more general tokamak geometries \cite{Bernard_2022, Bernard_2023, Bernard_2024}, and has recently been extended to model axisymmetric tokamak configurations similar to 2D fluid codes \cite{Shukla_2025}. A major challenge to simulate the SOL with high collisionality is that all Fokker-Planck based operators, including the general Landau or Rosenbluth forms \cite{Rosenbluth_1957} and reduced forms such as the Lenard-Bernstein-Dougherty (LBD) \cite{Lenard_1958, Dougherty_1964}, have stiff parabolic (i.e., diffusion) terms in velocity space. Explicit time integration of this term suffers from tiny time steps and makes the simulation very expensive. One natural thought is to switch to an implicit time integration scheme, since implicit time stepping schemes can be unconditionally stable and allow for arbitrarily large time steps. However, the nonlinear integro-differential form of the Fokker-Planck collision operators makes the implementation of fully implicit algorithms difficult but it can be done \cite{Taitano_2018, Killeen_1986, Ghosh_2018}. 

Therefore, we consider a simplified approximation to the Fokker-Planck collision operator, the Bhatnagar–Gross–Krook (BGK) collision operator \cite{Bhatnagar_1954}. While it misses some of the details of the full collision operator that can be important in some cases, it reproduces key features of collisions, such as causing the distribution function to relax towards a Maxwellian and preserving key conservation laws for particles, momentum, and energy. It can also match the momentum and energy exchange rates between species. Taking advantage of the simple form of this operator, we have implemented an implicit BGK collision operator in the framework of Gkeyll in order to relieve the stringent restriction of time steps and achieve gyrokinetic simulations in the high collisionality regime. 

Both the standard BGK and LBD models assume that the collision frequency is independent of velocity, missing the effect that high velocity particles are less collisional in a plasma.  While one can adjust coefficients to match some processes \cite{Rosen_2025}, it is not possible to exactly match multiple transport coefficients simultaneously (parallel and perpendicular thermal diffusivity, viscosity, resistivity, etc.) and there can be factors of 2-4 differences in some parameters. However, one can generalize these models to keep velocity dependence in the collision frequency\cite{Struchtrup_1997, Mieussens_2004, Haack_2021, Haack_2023}, which can do a much better job of matching various transport coefficients simultaneously, including the thermal force (related to the Soret effect). For example, the standard BGK model predicts a Prandtl number (ratio of momentum diffusivity to thermal diffusivity) of ${\rm Pr} = 1$ while a velocity-dependent BGK model can reproduce the proper value of ${\rm Pr} = 2/3$ for a monatomic gas\cite{Mieussens_2004}. 

At low collision frequency, the velocity diffusion in a Fokker-Planck type of collision operator enhances the impact of collisions on small scales in velocity (such as in narrow trapping regions). This effect is missing from a BGK model, which is better at treating large scale features. A simple way to improve this in future work would be to use a hybrid method, which uses an explicit LBD or explicit full Rosenbluth Fokker-Planck operator in most of the plasma, where the collision frequency is low, while transitioning to an implicit treatment of the BGK operator (as we discuss here) in colder regions of the plasma, where the collision frequency is high. Various collision operators have been used in recent gyrokinetic studies \cite{Belli_2012, Dorf_2014, Knyazev_2023, Ulbl_2023, Hoffmann_Frei_Ricci_2023}. 

The principal challenge in implementing a BGK operator is conservation of the momentum and energy, which is a fundamental requirement for collision operators. Gkeyll employs a discontinuous Galerkin (DG) method \cite{Cockburn_1998, Hakim_2020, Francisquez_2020} which combines some desirable features of finite volume methods and finite element methods. DG discretization of the equilibrium Maxwellian distribution to which the BGK relaxes introduces numerical errors and breaks the conservation. We have applied an iterative scheme to correct the discretized Maxwellian such that the BGK operator is conservative. This scheme is capable of handling the nonlinearity of the Maxwellian in momentum and energy. Then it is straightforward to generalize an explicit-implicit (IMEX) method \cite{Johnson_2025} to the gyrokinetic equation. By treating the collisionless advection term explicitly and the collision term implicitly, the restriction of time steps comes only from the collisionless advection term, which leads to significant speed-up for simulations at high collisionality. 

With the new capabilities provided by the implicit BGK, we have performed a number of 2D axisymmetric simulations of ASDEX-Upgrade (AUG) SOL using Gkeyll. In this work, we demonstrate that the parallel profiles of density, temperature, and electric potential produced by simulations with the implicit BGK closely resemble those obtained from explicit LBD simulations. However, the implicit BGK offers the advantage of being able to take larger time steps and work with lower phase space resolution, both of which contribute to a reduction of computational cost without loss of accuracy from the simplified form of the BGK operator. 

This paper is structured as follows. In Sect.~\ref{sec: implement}, we describe the procedures of implementing a conservative, implicit BGK collision operator and its cross-species generalization in the Gkeyll code. In Sect.~\ref{sec: benchmark}, we show the results of the benchmark tests for the features of the implicit BGK collision operator. In Sect.~\ref{sec: axisymmetric sims}, we show axisymmetric simulations of the ASDEX-Upgrade SOL and the comparisons between the implicit BGK and the explicit LBD operators. Finally, we summarize the results of this study and look ahead to future work in Sect.~\ref{sec: conlusions}.

\section{\label{sec: implement}The implicit BGK collision operator}
In the plasma kinetic equation, collision effects are described by a Fokker-Planck operator (FPO) which drives the plasma to thermal equilibrium through drag and diffusion in velocity space while conserving particle number, momentum, and energy:
\begin{equation}
    \cC[f_s] = -\nabla_{\bv}\cdot(\ba f_s - \bD\cdot\nabla_{\bv}f_s),
\end{equation}
where $f_s$ is the distribution function of species $s$, $\ba$ is the acceleration vector, and $\bD$ is the tensor diffusion coefficient.

A simplification sometimes used for this full Fokker-Planck operator is the LBD operator\cite{Lenard_1958, Dougherty_1964}, which maintains the Fokker-Planck advection-diffusion structure, with simplified drag and diffusion coefficients. The gyroaveraged form of the LBD operator is given by \cite{Francisquez_2022, Ulbl_2021}
\begin{alignat}{2}\label{eq: lbo}
    \cC[f_s] = \sum_r \nu_{sr} & \left\{ \pdv{}{v_{\parallel}}\left[(v_{\parallel}-u_{\parallel sr})f_s+v_{tsr}^2\pdv{f_s}{v_{\parallel}}\right] \right. \nonumber \\
    & \left. + \pdv{}{\mu}\left[2\mu f_s+2\frac{m_sv_{tsr}^2}{B}\pdv{f_s}{\mu}\right] \right\}.
\end{alignat}
The sum over $r$ denotes collisions between species $s$ and all the other species. $\nu_{sr}$ is the collision frequency between species $s$ and $r$. For self-species collisions (i.e., $s=r$),  $u_{\parallel ss}=u_{\parallel s}$ and $v_{tss}^2=v_{ts}^2$ are the mean parallel velocity and the squared thermal velocity of species $s$ as defined in Eq.~\ref{eq: definition of moments}.

An even more drastic simplification is to use a Bhatnagar–Gross–Krook (BGK) collision operator \cite{Bhatnagar_1954, Greene_1973} of the form
\begin{equation}\label{eq: bgk}
    \cC[f_s] = -\sum_r \nu_{sr}(f_s-f_{Msr}).
\end{equation}
The BGK operator relaxes the distribution function towards a Maxwellian $f_M$ at a certain collision frequency. For self-species collisions, $f_{Mss}=f_{Ms}$. Small-angle scattering is neglected here, so the operator is inaccurate for the relaxation of sharp gradients in velocity space.
 
In the general case when utilizing a Fokker-Planck operator for collisions, the gyrokinetic equation is a mixed hyperbolic-parabolic PDE, with the parabolic piece arising due to velocity space diffusion. Thus, the maximum stable explicit time step for the parabolic (collision) term $\Delta t_P$ can be much smaller than that for the hyperbolic (collisionless advection) term $\Delta t_H$ at high collision frequency and high velocity space resolution. Our previous explicit method with LBD suffers from overly restrictive $\Delta t_P$, which increases the simulation time. To achieve a high collisionality regime with reasonable computational cost, the time step limit from the collision term must be relieved. 

An implicit time stepping scheme formally allows a time step independent of collision frequency. However, as is shown in Eq.~\ref{eq: lbo}, the LBD model includes first and second order $v_\parallel$ and $\mu$ derivatives of the distribution function. The inversion of the LBD collision operator, which would be required by the implicit time integration scheme, can be computationally costly and difficult to parallelize. Implicit methods for the LBD operator, and even for the full Rosenbluth collision operator have been implemented in codes \cite{Killeen_1986}, including with an efficient multigrid method \cite{Taitano_2018, Ghosh_2018}. While they allow much larger time steps, they can be complicated to implement and have a computational cost per time step that is much larger than for the simple implicit BKG treatment we use here. An alternative time integration scheme for the LBD model is the super-time stepping (STS) method \cite{Meyer_2014}, which utilizes Legendre polynomials and chooses the internal stages in the explicit RK methods intelligently such that the overall time step is stable up to $\sim s^2\Delta t_P$ (s is the number of stages). In the context of AUG experiments, $\Delta t_H$ is approximately $\cO(10^2)$ times larger than $\Delta t_P$. Applying STS to LBD will lead to a speed-up with a factor of $\sqrt{\Delta t_H/\Delta t_P}\sim\cO(10)$, which is much smaller than that of the implicit BGK, $\Delta t_H/\Delta t_P\sim\cO(10^2)$. 

By contrast, besides removing the time step limit from the collision term, our implicit treatment of a BGK operator is relatively simple as well. Only a few iterations are required to preserve the conservation properties, and the implicit BGK operator works on even relatively low-resolution velocity grids. An implicit BGK, say with a backward Euler method, would evaluate both terms on the right-hand side of equation~\ref{eq: bgk} at the new time step; but if the discrete Maxwellian $f_{Ms}$ can be created with the exact same moments as $f_s$, only the first term needs to be evaluated at the new time step. This simpler approach necessitates the ability to assemble a discrete Maxwellian that preserves the moments of $f_s$, which we address next.

\subsection{\label{subsec: conservation}Conservation of moments}
In Eq.\ref{eq: bgk}, $f_{Ms}$ represents the Maxwellian distribution function of species $s$. The gyrokinetic form of $f_{Ms}$ is given by
\begin{equation}\label{eq: maxwellian}
    f_{Ms} = \frac{n_s}{(2\pi v_{ts}^2)^{d_v/2}}\exp\left[-\frac{(v_{\parallel}-u_{\parallel s})^2+2\mu B_0/m_s}{2v_{ts}^2}\right],
\end{equation}
where $n_s$, $u_{\parallel s}$, and $v_{ts}^2$ are the density, mean parallel velocity, and thermal velocity squared of species s, and $d_v$ signifies the number of physical velocity-space dimensions represented (some test problems for the code use $d_v=1$, but all of the cases in this paper are done with $d_v=3$). These quantities are calculated with different velocity moments of the distribution function
\begin{subequations}\label{eq: definition of moments}
    \begin{equation}
        n_s = \int f_s d\bv
    \end{equation}
    \begin{equation}
        u_{\parallel s} = \frac{1}{n_s} \int v_{\parallel}f_s d\bv
    \end{equation}
        \begin{equation}
        v_{ts}^2 = \frac{1}{d_v n_s}\int \left[ (v_\parallel - u_{\parallel s})^2 + \frac{2 \mu B}{m_s} \right] f_s d\bv 
    \end{equation}
\end{subequations}
where $d\bv=(2\pi B/m)dv_{\parallel}d\mu$, and we adopt the convention $T_s=m_sv_{ts}^2$. For simplicity, we write them as a vector $\bM_s=(n_s, u_{\parallel s}, v_{ts}^2)$, and the subscript $s$ is dropped in the rest of this subsection. To represent the Maxwellian numerically, $f_{M}$ is then projected onto the DG basis functions using Gauss-Legendre quadrature \cite{HakimJuno_2020}
\begin{equation}
    f_M(\bv) = \sum_p f_p\psi_p (\mathbf{\eta}(\bv)),
\end{equation}
where $\psi_p$ is the orthonormal polynomial basis function of $p$th order, and $\mathbf{\eta}(\bv)$ maps the physical space to logical space. However, this projection introduces discretization errors\cite{Johnson_2025}. 

Let $f_M(\bM)$ be the numerical approximation of the Maxwellian in Eq.~\ref{eq: maxwellian} based on a set of moments $\bM$, and let $\widehat{M}[f]$ be an operator to calculate the velocity moments of a distribution function $f$. In general, the moments of the numerical Maxwellian do not exactly match the input moments used to construct it, i.e., $\widehat{M}[f_{M}(\bM)]\neq \bM$. To ensure the moments are conserved in the BGK operator, we need to correct the projected Maxwellian, i.e., we need to find $\bM_{\rm in}$ such that $\widehat{M}[f_M(\bM_{\rm in})] = \widehat{M}[f]$. Since the Maxwellian is linear in the density, the density can be corrected by rescaling. However, $f_{M}$ is nonlinear in the drift velocity and thermal velocity squared, so we need to correct $u_{\parallel}$ and $v_t^2$ using an iteration method \cite{Mieussens_2000}. To correct the projected Maxwellian, we use a fixed point iteration scheme, of the form: 
\begin{subequations}\label{eq: iteration}
    \begin{equation}
        \Delta \bM^{(k)} = \widehat{M}[f] -\widehat{M}\left[f_M(\bM^{(k)})\right],
    \end{equation}
    \begin{equation}
        \bM^{(k+1)} = \bM^{(k)} + \Delta \bM^{(k)}.
    \end{equation}
\end{subequations}
Here $k$ in the superscript represents the iteration number, and we take $\bM^{(0)}=\widehat{M}[f]$. $f_M(\bM^{(k)})$ is the numerical Maxwellian constructed with the moments $\bM^{(k)}$. Note that $\bM$ is a vector representing the expansion of the moments $v_t^2(x)$ and $u_{\parallel}(x)$ in basis functions within each cell.  We use a stopping criterion based only on the accuracy of the cell averages of the moments, i.e., we iterate until the cell-average of $\widehat{M}[f_M(\bM_{\rm in})] = \widehat{M}[f]$ agrees in each cell to a relative accuracy of $10^{-12}$ relative to the reference thermal velocity $v_{t0}$ and thermal velocity squared $v_{t0}^2$, ensuring that global momentum and energy conservation reaches a desired level of accuracy. 

\subsection{\label{subsec: implicit}Implicit time stepping}
Since the major time step restriction comes from the collision term in the regime of interest, we will apply an implicit-explicit (IMEX) scheme that keeps the collisionless terms explicit as before, while treating the collision terms implicitly. With a first order split (sometimes referred to as a Godunov split), the gyrokinetic equation 
\begin{equation}\label{eq: gk}
    \pdv{\cJ f_s}{t} + \nabla \cdot \left(\cJ\dot{\bR}f_s\right) + \pdv*{\left(\cJ \dot{v}_{\parallel}f_s\right)}{v_{\parallel}}= \cJ C[f_s],
\end{equation}
is decomposed into a collisionless advection updater and a collision updater of the distribution function
\begin{subequations}
    \begin{alignat}{2}
        \label{eq: collisionless term}
            \pdv{\cJ f_s}{t} & = - \nabla \cdot \left(\cJ\dot{\bR}f_s\right) - \pdv*{\left(\cJ \dot{v}_{\parallel}f_s\right)}{v_{\parallel}}, 
            \\
        \label{eq: collision term}
            \pdv{\cJ f_s}{t} & = \cJ C[f_s],
    \end{alignat}
\end{subequations}
where $\cJ$ is the Jacobian of the gyrocenter coordinates, and $\dot{\bR}$ and $\dot{v}_{\parallel}$ are the phase space advection velocities \cite{Francisquez_2025}. This approach can be taken with any operator splitting technique, e.g., Strang splitting \cite{Strang_1968}. At the $n$th step, $f_s^{(n)}$ is first updated to $f_s^*$, the distribution function of an intermediate step, by the collisionless term as shown in Eq.\ref{eq: collisionless term}, using the original Strong Stability Preserving \cite{Shu_2002} (SSP) third-order Runge-Kutta (RK3) explicit method. And we can obtain the moments of $f^*$
\begin{equation}
    \bM_s^* = \widehat{M}[f_s^*].
\end{equation}
Then we update $f_s^*$ with the collision term as shown in Eq.\ref{eq: collision term} using the backward Euler method
\begin{equation}
    \frac{f_s^{(n+1)}-f_s^*}{\Delta t} = -\nu_{ss}\left[f_s^{(n+1)}-f_{Ms}(\bM_{s,in}^*)\right] 
    \label{eq:partial-imp-bgk}
\end{equation}
to obtain an expression of the distribution function at the $(n+1)$th step
\begin{equation}
    f_s^{(n+1)}=\frac{f_s^*+\nu_{ss}\Delta t f_{Ms}(\bM_{s,in}^*)}{1+\nu_{ss}\Delta t}.
\end{equation}
Note that $f_{Ms}$ is projected with a set of moments $\bM_{s,in}^*$ obtained from the iteration described in Eq.~\ref{eq: iteration} such that $\widehat{M}[f_M(\bM_{s,in}^*)] = \widehat{M}[f_s^*]$. Now that the restriction of the time step only comes from the collisionless terms, we can take larger time steps to speed up the simulations. Here we have chosen Godunov splitting with a first-order backward Euler method for the BGK operator for simplicity and because it is robust, being positivity preserving and L-stable ($f_s^{(n+1)}$ reduces to a Maxwellian in the limit of $\nu_{ss} \Delta t \rightarrow \infty$). A Godunov splitting can be made second order accurate like Strang splitting if the method for each term is second order accurate \cite{LeVeque_2002}. There are several ways this could be done, and exploring these we leave for future work. One is to replace $\nu_{ss} \Delta t$ in these formulas with ($\nu_{ss} \Delta t + (\nu_{ss} \Delta t)^2/2)$, the next order expansion of the exact result $\exp(\nu_{ss} \Delta t) - 1$. Related topics are discussed in the conclusion section.

\subsection{\label{subsec: multispecies}Multispecies BGK}
To generalize the BGK operator to collisions between different species $s$ and $r$, we must determine the cross Maxwellian $f_{Msr}$ which appears in Eq.\ref{eq: bgk}
\begin{equation}
    f_{Msr} = \frac{n_{sr}}{(2\pi v_{tsr}^2)^{d_v/2}}\mathrm{exp}\left[-\frac{(v_{\parallel}-u_{\parallel sr})^2+2\mu B_0/m_s}{2v_{tsr}^2}\right].
\end{equation}
Here, $\bu_{sr}$ and $v_{tsr}$ are intermediate flow and thermal velocities which govern how the multispecies plasma equilibrates. The calculation of the cross Maxwellian for BGK follows that for LBD \cite{Francisquez_2022}; specifically, we use the Greene's form \cite{Greene_1973} of the cross-species Maxwellian (i.e., the BGK-G in the terminology of \cite{Haack_2017}). We can see clearly that $n_{sr}=n_s$, since the number of particles of each species stays the same before and after the collision. To determine the cross mean parallel velocity and the thermal velocity for the two colliding species, the conditions we have \cite{Morse_1963, Greene_1973} are the conservation of momentum and energy
\begin{subequations}
    \begin{alignat}{2}
        \int m_sv_{\parallel}C[f_s,f_r] d\bv + \int m_rv_{\parallel}C[f_r,f_s] d\bv &= 0, \\
        \int \frac{1}{2}m_sv^2C[f_s,f_r] d\bv + \int \frac{1}{2}m_rv^2C[f_r,f_s] d\bv &= 0,
    \end{alignat}
\end{subequations}
and the Morse relaxation rate of momentum and thermal energy (for Coulomb collisions between two Maxwellian distributions)
\begin{widetext}
    \begin{subequations}
    \begin{alignat}{2}
        \pdv*{(m_sn_su_{\parallel s}-m_rn_ru_{\parallel r})}{t} &= -\alpha_E(m_s+m_r)(u_{\parallel s}-u_{\parallel r}), \\
        \pdv*{\left(\frac{d_v}{2}m_sn_sv_{ts}^2-\frac{d_v}{2}m_rn_rv_{tr}^2\right)}{t} &= -\alpha_E\left[d_v(m_sv_{ts}^2-m_rv_{tr}^2)+\frac{1}{2}(m_s-m_r)(u_{\parallel s}-u_{\parallel r})^2\right].
    \end{alignat}
    \end{subequations}
\end{widetext}
Here $\alpha_E$ is the relaxation coefficient defined by Morse \cite{Morse_1963} as
\begin{equation}
    \alpha_E = \frac{2n_sn_r(q_sq_r)^2\mathrm{log}\Lambda_{sr}}{3(2\pi)^{3/2} \varepsilon_0^2 m_s m_r(v_{ts}^2+v_{tr}^2)^{3/2}},
\end{equation}
where $\mathrm{log}\Lambda_{sr}$ is the Coulomb logarithm. 

Taking momentum and energy moments of the specific form of the cross species BGK operator
\begin{equation}
    \left.\pdv{f_s}{t}\right|_c = C[f_s,f_r] = \nu_{sr}(f_{Msr}-f_s),
    \label{eq:cross-spec-BGK}
\end{equation}
and requiring that the momentum and energy exchange rates match the Morse results partially constrains the choice of $\nu_{sr}$, $u_{sr}$, and $v_{tsr}^2$ 
(and $\nu_{rs}$, $u_{rs}$, and $v_{trs}^2$), but there is a free parameter left.  
Greene introduced a free parameter, $\beta$, which parametrizes the choice of $u_{sr}$ as a weighted mean between $u_s$ and $u_r$.
We slightly generalize Greene's Eq.~14 to
\begin{align}
        u_{\parallel sr} = & \, u_{\parallel s} + \frac{(1+\beta)\delta_{sr}}{2}\left(u_{\parallel r}-u_{\parallel s}\right),
\end{align}
to allow this weighting to depend on the species $s$ and $r$ parameters if desired.
The parameters $\beta$ and $\delta_{sr}$ always appear in the combination $(1 + \beta)\delta_{sr}$ and so aren't independent, but we leave them like this to be able to more easily compare with various past choices of BGK-type operators.

Matching the Morse momentum and energy exchange rates then constrains the cross-species temperature to be
\begin{align} 
\label{eq: cross prim moms}
        v_{tsr}^2 = & \, v_{ts}^2 + \frac{(1+\beta)\delta_{sr}}{m_s+m_r}\left(m_rv_{tr}^2-m_sv_{ts}^2\right)+ \nonumber \\
        & \frac{1}{d_v}\left[\frac{m_r(1+\beta)\delta_{sr}}{m_s+m_r} - \frac{(1+\beta)^2\delta_{sr}^2}{4}\right]\left(u_{\parallel r}-u_{\parallel s}\right)^2, 
\end{align}
and constrains the collision frequency to be
\begin{equation}
    \nu_{sr} = \frac{\alpha_E (m_s+m_r)}{(1+\beta)\delta_{sr} m_s n_s}.
\end{equation}
Note that this gives $\nu_{sr} / \nu_{rs} = \delta_{rs} m_r n_r / (\delta_{sr} m_s n_s)$, so Greene's assumption that $\nu_{sr} m_s n_s = \nu_{rs} m_r n_r$ required $\delta_{sr} = \delta_{rs}$, but one does not need to make that assumption.
One could choose $\delta_{sr} (1+\beta)$ from some other property, such as the pitch-angle-scattering isotropization rate.  
For the results in this paper, we set $\beta = 0$ and $\delta_{sr} = 1$, except for self-species collisions where $\delta_{ss} = 2$.

In the following simulations, we will use a spatially varying cross collision frequency that takes the evolution of density and thermal velocity into consideration
\begin{equation}
    \nu_{sr}(x,t) = \nu_{sr0}\frac{n_r(x,t)}{n_{r0}}\frac{(v_{ts0}^2+v_{tr0}^2)^{3/2}}{(v_{ts}^2(x,t)+v_{tr}^2(x,t))^{3/2}}, 
\end{equation}
where the subscript $0$ indicates the reference values of density and thermal velocity. At present we use this to find the variation of $\nu_{sr}$ between DG cells, but it is taken as a constant within a cell.

For self-collisions, the momentum and energy moments are conserved during a collision time step, so in Eq.\ref{eq:partial-imp-bgk}, 
$f_s^*$ and $f_s^{(n+1)}$ have the same moments $\bM_s^*$, so $f_{Ms}({\bM}^*_s)$ on the RHS of Eq.\ref{eq:partial-imp-bgk} does not change during the time step and can be treated explicitly.
This is not rigorous for cross-species collisions, so a fully implicit treatment of Eq.\ref{eq:cross-spec-BGK} would require evaluating $f_{Msr}$ at the future time level $n+1$ in order to exactly conserve total energy and momentum. This can be done at the small cost of solving two $N_s \times N_s$ matrix problems (where $N_s$ is the number of species) to find $u_{\parallel s}^{(n+1)}$ and $T_s^{(n+1)}$ at the future time level. The errors vanish for steady-state applications (such as the AUG test case done here).  For turbulence calculations, we will usually have $\nu_{\parallel sr} \Delta t \ll 1$ except for electron-electron and electron-ion collisions (which are larger than $\nu_{ie}$ by a factor of $m_i/m_e$). For electron-ion collisions, the recoil of ions to conserve energy and momentum is very small and can be neglected in most cases. This partially-implicit method still gives the main advantage of being able to take large stable time steps, and still gives good results for quantities that vary slowly on the time scale of the time step $\Delta t$.

As shown in Eq.~\ref{eq: cross prim moms}, there is a term with negative sign in the expression of $v_{tsr}^2$ calculated from Greene's formula \cite{Greene_1973}. In simulations with realistic parameters, we have $m_i\gg m_e$. When the relative flow $|u_{\parallel i}-u_{\parallel e}|$ becomes supersonic, the cross thermal velocity squared for ion-electron collision $v_{tie}^2$ can get negative, which is not physical. Currently we turn off the cross species collision in a cell if the cross thermal velocity squared is negative at any of the Gauss-Legendre quadrature points (this is rare). A modified form of the BGK operator could improve the handling of this, such as limiting the cross-species energy exchange rate in the rare cases where the relative drift is supersonic. (This would still preserve isotropization and the resulting resistivity, which sometimes are more important.) Also, it may be better to enforce realizability of the BGK parameters at positivity control points instead of the Gaussian quadrature points in a cell \cite{Mandell_thesis}. The linear expansion of a realizable distribution function must be positive at the positivity control points while it could be negative at the Gaussian quadrature points, making the code more robust at low temperature.

\section{\label{sec: benchmark}Benchmarks of implicit BGK}

The implicit BGK described in Sec.~\ref{sec: implement} has been implemented in the DG gyrokinetic solver of the Gkeyll computational plasma physics framework \cite{Francisquez_2025}. Before applying it to simulations of tokamak experiments, we have carried out a series of tests in 1D slab geometry to demonstrate the properties of the implicit BGK and verify the implementation. Here we present three sets of tests: Sec.~\ref{subsec: conservation test} shows the performance of the iteration scheme on correcting the discretized Maxwellian to conserve moments; Sec.~\ref{subsec: implicit speed-up}
shows the speed-up achieved by the implicit time integration scheme of the collision term; Sec.~\ref{subsec: cross collisions} shows the validation of implicit BGK generalized to deal with multispecies collisions. The input files for all tests discussed in this section can be found in the GitHub repository \url{https://github.com/ammarhakim/gkyl-paper-inp/tree/master/2025_PoP_BGK}. 

\subsection{\label{subsec: conservation test}Conservation test}
We check that the density, momentum, and energy are indeed conserved during the evolution of a spatially uniform bump-on-tail distribution with the BGK collision operator in 1D configuration space. Here we only test the correction of the Maxwellian projection. For this test, we utilize lower collisionality and an explicit time stepping scheme. For simplicity, all the physical quantities are taken to be dimensionless. The mass $m=1.0$, the charge $q=1.0$, the reference magnetic field $B_0=1.0$, and the collision frequency $\nu=0.01$. The distribution is initialized as 
\begin{alignat}{2}\label{eq: bump}
    f(v_{\parallel}, \mu) = & \frac{n_0}{(2\pi v_{t0}^2)^{3/2}}\mathrm{exp}\left[-\frac{(v_{\parallel}-u_{\parallel 0})^2+\frac{2\mu B_0}{m}}{2v_{t0}^2}\right] \\
    & + \frac{n_b}{(2\pi )^{3/2}v_{t0}^2 v_{tb}^{}}\mathrm{exp}\left[-\frac{(v_{\parallel}-u_{\parallel b})^2}{2v_{tb}^2}-\frac{2\mu B_0}{2 m v_{t0}^2}\right], \nonumber
\end{alignat}
where the first term is a reference Maxwellian, with number density $n_0=1.0$, mean parallel velocity $u_{\parallel 0}=0.0$, and thermal velocity $v_{t0}=1.0$, and the second term represents the bump, with number density $n_b=0.25$, location in velocity space $u_{\parallel b}=2.0$, and thermal velocity $v_{tb}=0.3$. The simulation domain in the phase space is bounded by $[0.0, 1.0]$ in the $x$ direction, $[-6.0v_{t0}, 6.0v_{t0}]$ in the parallel velocity direction, and $[0.0, 16.0\mu_0]$ in the magnetic moment direction, where $\mu_0=mv_{t0}^2/(2.0B_0)$. The phase space of the simulation domain is discretized uniformly into $2\times32\times32$ cells. The basis functions are linear polynomials in $x$ and $\mu$ and quadratic polynomials in $v_{\parallel}$ \cite{Francisquez_2025}, which are used in all the following benchmark tests.

\begin{figure}[bh]
\centering
\includegraphics[width=\linewidth]{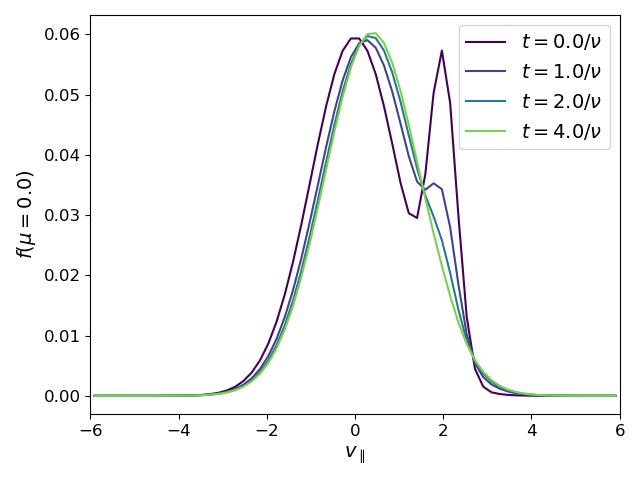}
\caption{\label{fig: bump relaxation}Relaxation of a bump-on-tail distribution. The distribution along the parallel velocity is shown at $\mu=0.0$ at four time slices.}
\end{figure}

\begin{figure}[bh]
\centering
\includegraphics[width=\linewidth]{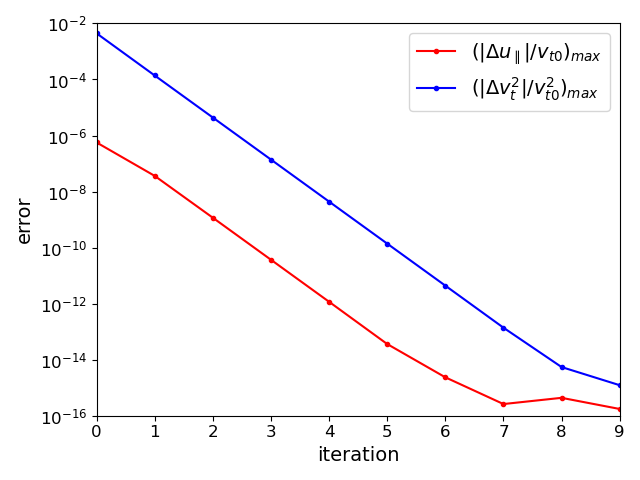}
\caption{\label{fig: error}The maximum relative errors of parallel velocity and error of thermal velocity squared in configuration space drop to near the machine precision within $9$ iterations.}
\end{figure}

This test is run for several collision times. Figure~\ref{fig: bump relaxation} shows the distribution function along parallel velocity at $\mu=0.0$ at $t=0.0, 1.0/\nu, 2.0/\nu, 4.0/\nu$. As time evolves, the bump gradually diminishes in prominence due to collisions. At $t=4.0/\nu$, the distribution has relaxed entirely to a Maxwellian. Figure~\ref{fig: error} shows the relative errors of the mean parallel velocity and the thermal velocity squared for the correction of the Maxwellian projection at the beginning of the first time step. Here the relative errors are the maximum among all the cells in the configuration space. The relative errors decreases rapidly with the number of iterations in Eq.~\ref{eq: iteration}, and these errors drop to near the machine precision within $9$ iterations. We thus argue that utilizing $\cO(10)$ iterations in production simulations is sufficient to reduce the discretization errors in the projected moments by a significant amount and in all subsequent simulations we set the maximum iteration count to 10. The number of iterations it takes for the correction to converge depends on the extent and the resolution of the velocity space \cite{Johnson_2025}, hence it is important to choose an appropriate velocity mesh that converges fairly rapidly while maintaining computational efficiency. 

\subsection{\label{subsec: implicit speed-up}Sod shock tests}
\begin{figure}[ht]
\centering
\includegraphics[width=\linewidth]{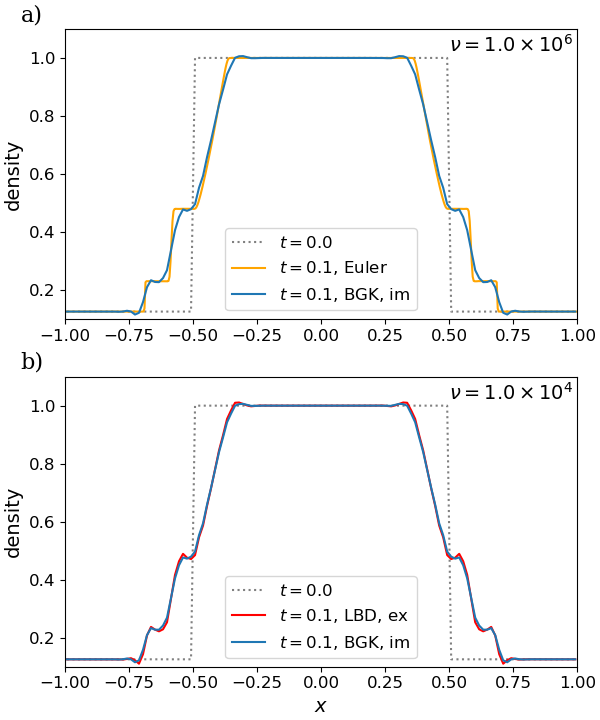}
\caption{\label{fig: sodshock}1D Sod shock propagation. (a) Comparison of solutions to Euler equations and to Gyrokinetic equation with implicit BGK. (b) Comparison of solutions to Gyrokinetic equation with implicit BGK and explicit LBD. The dotted lines represent initial density.}
\end{figure}

Next we demonstrate how much speed-up can be obtained from the implicit BGK operator compared to the explicit LBD operator by a Sod shock problem in 1D configuration space. For simplicity, all the physical quantities are rendered dimensionless. The mass $m=1.0$, the charge $q=1.0$, and the reference magnetic field $B_0=1.0$. As is shown by the dotted line in Fig.~\ref{fig: sodshock}, the density is inilialized as
\begin{equation}
    n(x) = \left\{ \begin{array}{ll}
         1.0, & -0.5<x<0.5 \\ 0.125,  & \mathrm{else} 
         \end{array}\right..
\end{equation}
The initial temperature profile is 
\begin{equation}
    T(x) = \left\{ \begin{array}{ll}
         1.0, & -0.5<x<0.5 \\ 0.8,  & \mathrm{else} 
         \end{array}\right..
\end{equation}
There is no initial flow in this system. The simulation domain in the phase space is bounded by $[-1.0, 1.0]$ in the $x$ direction, $[-6.0v_{t0}, 6.0v_{t0}]$ in the parallel velocity direction, and $[0.0, 18.0\mu_0]$ in the magnetic moment direction, where $v_{t0}=1.0$, $\mu_0=mv_{t0}^2/(2.0B_0)$. The phase space of the simulation domain is discretized uniformly into $64\times16\times16$ cells.

First, we verify that the solution of the gyrokinetic equation with the implicit BGK converges to the solution of the 5-moment fluid (Euler) equation in the fluid limit, i.e., at the limit of infinite collisionality. In the simulations, we use $\nu=1.0\times10^6$. In order to compare with the gyrokinetic solution of the implicit BGK test, we set $\gamma=5/3$ for monotonic gas with 3 velocity dimensions and use the same initial conditions in the Euler test. As is shown by Fig.~\ref{fig: sodshock}(a), the density profile of the implicit BGK test agrees well with that of the Euler test at $t=0.1$. Despite some minor overshoots, the characteristics of rarefaction waves, the contact discontinuities, and the shocks are captured in the implicit BGK tests. This demonstrates the sufficiency of the backward Euler method in terms of numerical accuracy. Then the results of the gyrokinetic simulations with the implicit BGK and the explicit LBD are compared with each other at $\nu=1.0\times10^4$. As is shown by plot (b) of Fig.~\ref{fig: sodshock}, the two simulations produce similar results. However, with all simulation settings the same, the implicit BGK time step is 20000 times larger than the explicit LBD time step, and implicit BGK case runs 7480 times faster than the explicit LBD case. 

\subsection{\label{subsec: cross collisions}Velocity and temperature relaxation tests}
Our final benchmark of the generalized implicit BGK is on the relaxation of the deuterium plasma to thermal equilibrium, following the examples of previous works on benchmarking different collision operators in different gyrokinetic codes\cite{Hager_2016, Ulbl_2021, Francisquez_2022}. In this anisotropic system of electrons and deuterium ions, the distributions are initialized as bi-Maxwellians
\begin{equation}
    f_s(v_{\parallel}, \mu)=\frac{n_0}{\alpha(2\pi v_{ts}^2)^{3/2}} \mathrm{exp}\left[-\frac{(v_{\parallel}-u_{\parallel s})^2+\frac{2\mu B}{(m_s\alpha)}}{2v_{ts}^2}\right]
\end{equation}
where $\alpha=1.3$, $B=1.0~\mathrm{T}$, $n_0=7.0\times10^{19}~\mathrm{m^{-3}}$. The initial temperatures are $T_{\parallel e0}=300.0~\mathrm{eV}$, $T_{\parallel i0}=200.0~\mathrm{eV}$, $T_{\perp s0}=\alpha T_{\parallel s0}$, and the reference temperatures are $T_{s0}=(2T_{\perp s0}+T_{\parallel s0})/3$. The reference thermal velocities are derived as $v_{ts0}=\sqrt{T_{s0}/m_s}$, and we set the reference mean parallel velocities to be $u_{\parallel i0}=50(m_e/m_i)v_{ti0}$, $u_{\parallel e0}=0.5\sqrt{m_e/m_i}v_{te0}$. The simulation domain in the phase space is bounded by $[-2.0, 2.0]~\mathrm{m}$ in the $x$ direction, $[-5.0v_{ts0},5.0v_{ts0}]$ in the parallel velocity direction, and $[0.0, 25.0\mu_{s0}]$ in the magnetic moment direction. This test is spatially uniform, and we only have $1$ cell in the configuration space. The domain in velocity space is discretized into $16\times16$ cells.

\begin{figure}
\includegraphics[width=\linewidth]{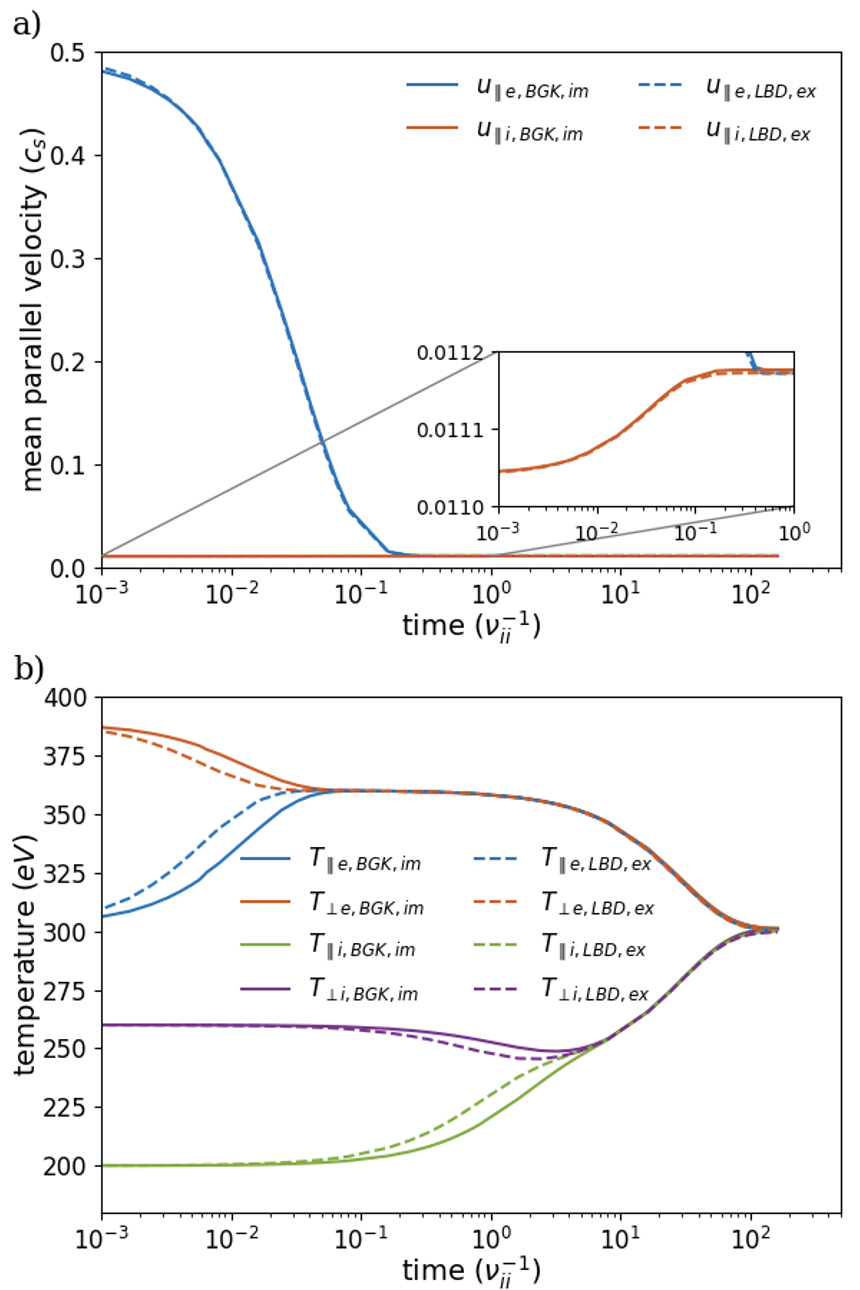}
\caption{\label{fig: cross_validation}Mean parallel velocity and temperature relaxation of an anisotropic deuterium plasma. The solid curves represent results of the test with the implicit BGK, and the dotted curves represent results of the test with the explicit LBD.}
\end{figure}

Figure~\ref{fig: cross_validation} shows the time evolution of the mean parallel velocities and the temperatures for the implicit BGK compared with the explicit LBD. The time axis is normalized to the ion collision time. As indicated in plot (a), both tests give similar results that the mean parallel velocities of electrons and ions equilibrate on the ion collision time scale at a velocity slightly larger than $u_{\parallel i0}$. As shown in plot (b), the electrons are isotropized first on the electron collision time scale. Then the ions are isotropized on the ion collision time scale. Finally, the ions and electrons equilibrate with each other on the ion-electron collision time scale. The results of the implicit BGK case are in good agreement with that of the explicit LBD case, except that the isotropization of both species occurs slightly more slowly for the implicit BGK case. This difference arises from the fact that the BGK operator and the LBD operator have different isotropization rates for $T_{\parallel}$ and $T_{\perp}$. However, this relaxation occurs quickly on the time scales of interest, and as we will show in the next section, does not impact the final macroscopic profiles. 

\section{\label{sec: axisymmetric sims}Axisymmetric ASDEX-Upgrade SOL simulations}
With all the benchmark tests shown in the previous section, we now deploy the implicit BGK collision operator on a production calculation based on previous work exploring axisymmetric transport in tokamaks. In particular, we have conducted a number of 2D axisymmetric simulations of the AUG SOL comparing the implicit BGK and the explicit LBD. The input files for all the tests discussed in this section can be found in the GitHub repository \url{https://github.com/ammarhakim/gkyl-paper-inp/tree/master/2025_PoP_BGK/axisymmetric_AUG_SOL}.

\begin{figure}[ht]
\centering
\includegraphics[width=\linewidth]{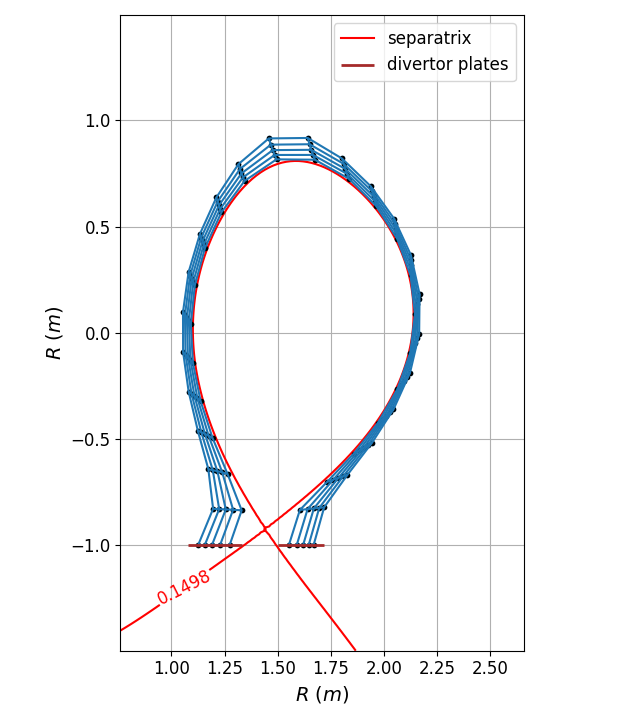}
\caption{\label{fig: nodes}Poloidal projection of the flux tube. The black dots represent the nodes; the blue lines represent the boundary of the cells. The divertor plates are horizontally placed for convenience. This plot only shows 4 cells in the x direction for clarity.}
\end{figure}

Overall, we utilize an axisymmetric setup similar to Francisquez \textit{et al.} \cite{Francisquez_2025} The magnetic geometry is generated from a numerical equilibrium taken from AUG, and the reference magnetic field is $B_0=2.57~\mathrm{T}$ at the major radius. Gkeyll uses a field-aligned coordinate system in the configuration space, with poloidal flux $\psi$ as the radial coordinate $x$, the normalized poloidal arc length
$\theta$ as the field following coordinate $z$, and $\alpha$ which satisfies $\bB=\nabla\alpha\times\nabla\psi$ as the field labeling coordinate $y$. Our 2D simulations do not consider variation in the $y$ direction and therefore are axisymmetric; in other words, we consider a simulation at fixed $\alpha$ for these axisymmetric simulations. For the lower single null geometry of AUG, our simulation domain is a flux tube in the AUG SOL extending from the outer divertor plate to the inner divertor plate, as shown in Fig.~\ref{fig: nodes}. Our flux tube lies entirely outside of the separatrix at $\psi=0.1498$. The flux tube is then mapped onto a rectangular computational box with the $x$ range $[0.154,0.170]$ and the $z$ range $[-\pi,\pi]$. The velocity space ranges from $[-6.0v_{ts}, 6.0v_{ts}]$ in the parallel velocity direction, and $[0.0, 16.0\mu_s]$ in the magnetic moment direction, where $v_{ts}=\sqrt{T_{s0}/m_s}$ and $\mu_s=T_{s0}/2B_0$ are calculated with the reference temperature $T_{s0}$. The $(x,z,v_{\parallel}, \mu)$ phase space of the simulation domain is discretized into $32\times24\times16\times8$ DG cells. This uses piecewise linear basis functions in all directions except $v_\parallel$, where piecewise quadratic basis functions are used, so this can be thought of as similar to a finite-difference grid of $64\times48\times48\times16$. The grid is uniform in the configuration space, but nonuniform in the velocity to resolve the larger dynamic range of temperature present in this calculation \cite{Francisquez_2025}.  

\begin{figure}[ht]
\centering
\includegraphics[width=\linewidth]{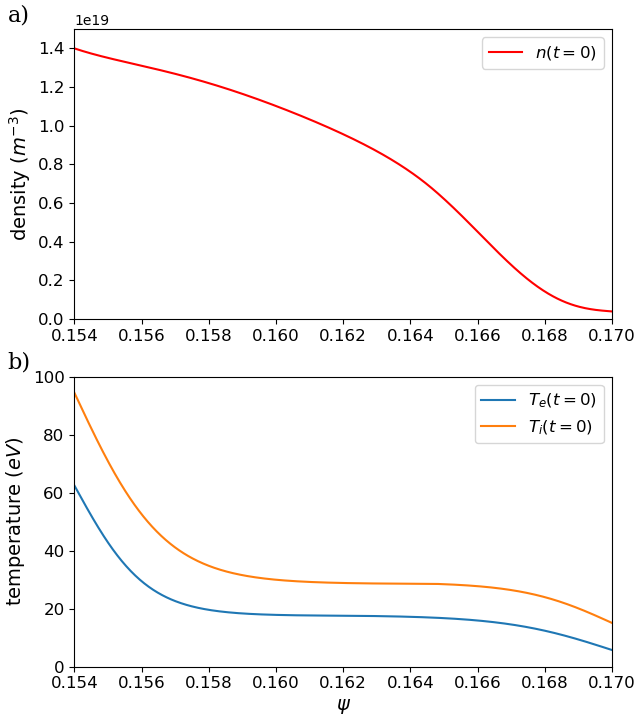}
\caption{\label{fig: ICs}Initial density and temperature profiles in the $x$ direction taken from the high effective collisionality case of EMC3-EIRENE simulations\cite{Carralero_2017, Lunt_2015}.}
\end{figure}

Our initial density and temperature profiles are functions of $x$ and uniform along $z$. The variations in the $x$ direction are taken from the the high effective collisionality case ($\Lambda_{div}>1$) of previous EMC3-EIRENE simulations\cite{Carralero_2017, Lunt_2015}. We fit the data with elementary functions and input them as initial conditions. The fitted curves are shown in Fig.~\ref{fig: ICs}. The density profile is relatively flat in a wide region on the lower radial side, and it becomes steep in the far SOL. The temperature profiles are steep near the separatrix, and relatively flat away from the separatrix. The reference density and temperatures $n_0=1.4\times10^{19}~\mathrm{m^{-3}}$, $T_{e0}=62.5~\mathrm{eV}$, and $T_{i0}=94.5~\mathrm{eV}$ are taken at the lower $x$ boundary.

For more accurate quantitative comparisons with experiments we will eventually need to include atomic physics including neutrals and radiation. We have done some work on these topics\cite{Bernard_2022, Shukla_2025, Roeltgen_2025} and plan more detailed atomic physics models in the future. Our main purpose here is to exercise the new implicit BGK collision operator and the gyrokinetic part of the code, so here we use a simplified source model that roughly mimics some of the missing physical effects. We inject a narrow Maxwellian source of particles near the inner boundary of the simulation to approximate the source of particles and energy that is being transported by turbulence from the confined part of the plasma into the SOL. The temperature of the main part of the source is set to be the observed temperature near the inner $x$ boundary (near the separatrix). The particle source rate is chosen so that the integrated power in this source is (approximately) the net heating in this experiment. This mimics some average of higher energy particles escaping from the plasma core into the SOL and ionization of colder neutrals in the plasma edge from recycling or gas injection. The main free parameter here is the temperature of the source. The resulting density and temperature profiles calculated by the code are found to be much broader than the source, so the details of the source shape do not matter very much.

The particle source is localized around the outboard midplane (OMP) to model the turbulent transport of plasma from the confined region into the SOL. The source is a Gaussian centered at $x_s=0.1574$ with a standard deviation $\lambda_s=0.0011$ in the radial direction, and it is cut off at $\pm L_z/8$ away from the OMP. The radial source center $x_s$ is a few gyroradii away from the separatrix such that it is more than a banana width away from the separatrix (the approximate inner boundary of our simulation) because we do not want a large fraction of our source plasma to be lost through the inner boundary in one transit or bounce orbit.  (We are working on better boundary conditions and sources near the inner boundary.) The particle source function is given by
\begin{align}
    n_{src}(x,z) = \left\{ \begin{array}{ll}
         n_{src0} \mathrm{max}\left[\mathrm{exp}\left(-\frac{(x-x_{s})^2}{2\lambda_s^2}\right), 0.01\right], \\
         \hfill \mathrm{if} \left|z+\frac{L_z}{4}\right|<\frac{L_z}{8} \\ 0.01n_{src0},  \hfill \mathrm{else}
         \end{array}\right.
\end{align}
The temperatures of the sources are step functions in the radial direction, and they are also cut off at $\pm L_z/8$ away from the OMP. The source temperature functions are given by
\begin{subequations}
    \begin{equation}
        T_{e,src}(x,z) = \left\{ \begin{array}{ll}
         T_{e0}, \hspace{0.2em} x<x_s+3\lambda_s~\mathrm{and}~\left|z+\frac{L_z}{4}\right|<\frac{L_z}{8} \\ T_e(x_{max}, t=0), \hfill \mathrm{else}
         \end{array}\right.
    \end{equation}
    \begin{equation}
        T_{i,src}(x,z) = \left\{ \begin{array}{ll}
         T_{i0}, \hspace{0.2em} x<x_s+3\lambda_s~\mathrm{and}~\left|z+\frac{L_z}{4}\right|<\frac{L_z}{8} \\ T_i(x_{max}, t=0), \hfill \mathrm{else} 
         \end{array}\right.
    \end{equation}
\end{subequations}
The parameter $n_{src0}$ is adjusted so that the input power is in the typical range of AUG ohmic heating and electron cyclotron heating powers. In this series of simulations we use $n_{src0}=24.0\times10^{22}~\mathrm{m^{-3}s^{-1}}$.

These 2D axisymmetric simulations do not have turbulence; therefore we add a diffusion term $\nabla\cdot(\bD\cdot\nabla f_s)$ to the right hand side of the gyrokinetic equation to account for turbulent cross-field transport (like SOLPS, UEDGE, and other axisymmetric fluid codes do). Here we set the radial particle diffusivity $D_{\perp}=0.1$.

Our flux tube starts and ends on divertor plates in the parallel direction. We apply a conducting sheath boundary condition model \cite{Shi_2019} for the particles at both the lower and upper $z$ boundaries so that electrons with energy less than $e\Phi_{sh}$ are reflected back into the simulation domain, where $\Phi_{sh}$ is the sheath potential. No boundary condition is required for the potential at the $z$ boundaries, since there are no derivatives of $z$ in the gyrokinetic Poisson equation \cite{Francisquez_2025}. 
At the inner $x$ boundary, We set the incoming distribution function to be Maxwellian with a density $n_e=n_i=1.4\times10^{19}\mathrm{m^{-3}}$, $T_e=62.5\mathrm{eV}$, and $T_i=94.5\mathrm{eV}$ (as in Fig.~\ref{fig: ICs}). This tends to offset related losses of particles at the inner $x$ boundary due to the $\nabla B$ and curvature drifts. The outer $x$ boundary is close to the first wall, so we let the particles get absorbed there. The electric potential is fixed to zero at both $x$ boundaries. Future work will consider improved boundary conditions for the potential.

\begin{figure*}
\centering
\includegraphics[width=\linewidth]{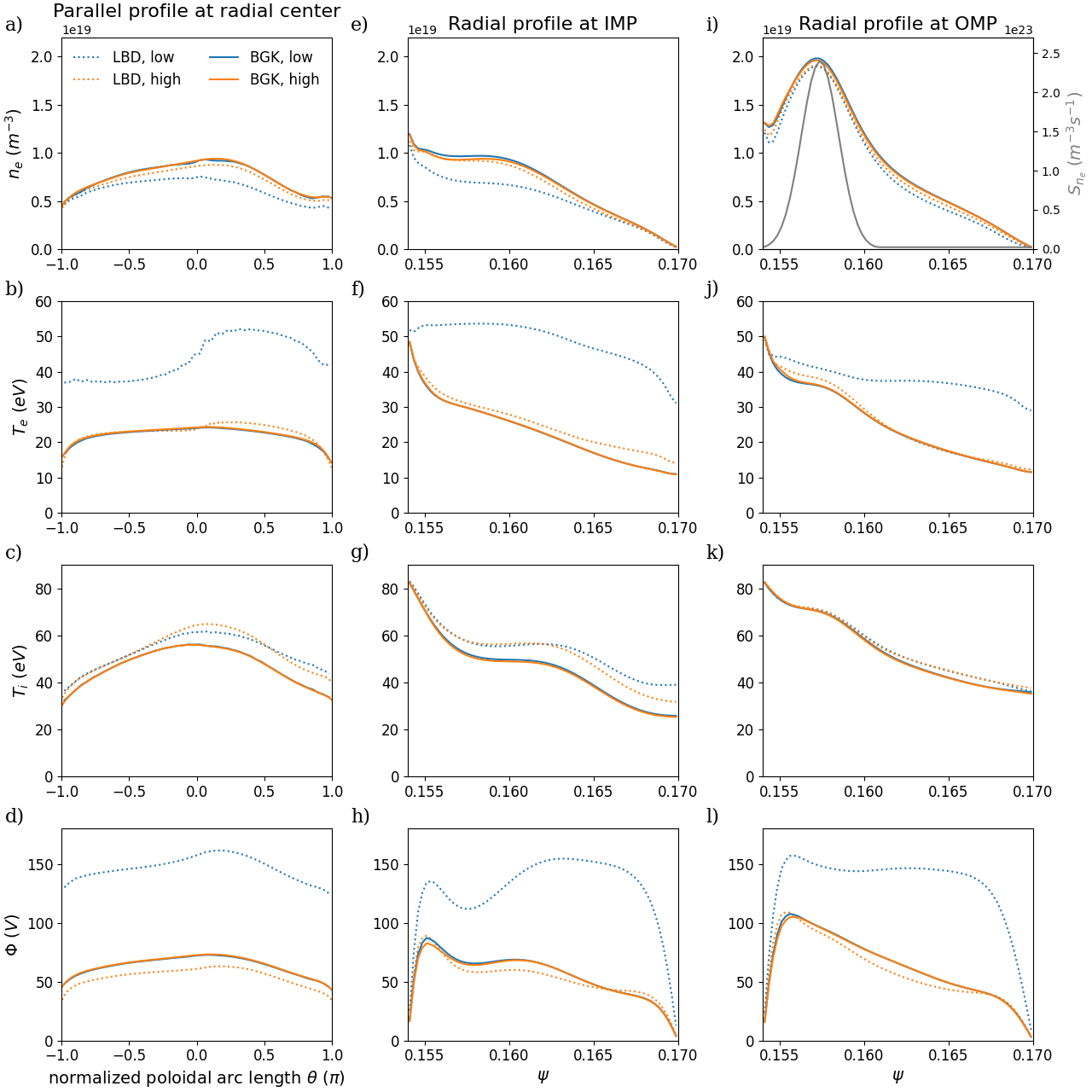}
\caption{\label{fig: profiles}Comparison of electron density, electron temperature, ion temperature, and electric field profiles for 2D axisymmetric AUG SOL simulation with the implicit BGK and the explicit LBD at two different phase space resolutions. (a)-(d) Parallel profiles at the radial domain center; (e)-(h) radial profiles at the inboard midplane (IMP); (i)-(l) radial profiles at the outboard midplane (OMP). The shape of the density source profile is indicated by the gray curve in panel (i), demonstrating that the resulting density profile is broader than the source. The solid curves are for the implicit BGK; the dotted curves are for the explicit LBD. The blue curves are for $(N_x, N_z, N_{v_{\parallel}},N_{\mu})=(32, 24, 16, 8)$; the orange curves are for $(N_x, N_z, N_{v_{\parallel}},N_{\mu})=(32, 96, 32, 16)$.}
\end{figure*} 

\begin{figure*}
\centering
\includegraphics[width=0.75\linewidth]{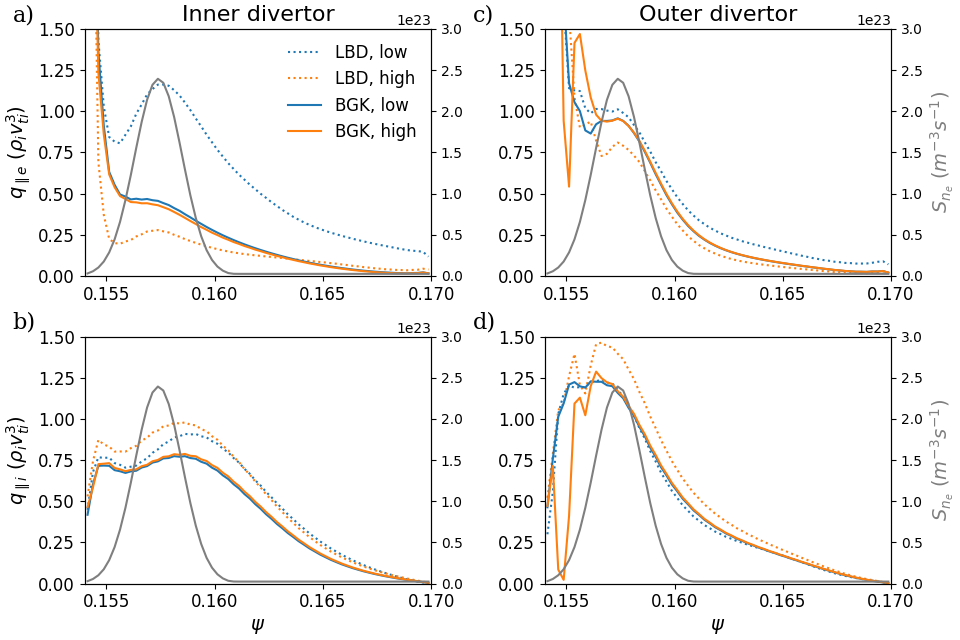}
\caption{\label{fig: heat flux profiles}Comparison of electron and ion parallel heat flux profiles for 2D axisymmetric AUG SOL simulation with the implicit BGK and the explicit LBD at two different phase space resolutions. (a)(b) Radial profiles at the inner divertor; (c)(d) radial profiles at the outer divertor, with the shape of the density source profile indicated by the gray curve. The solid curves are for the implicit BGK; the dotted curve is for the explicit LBD. The blue curves are for $(N_x, N_z, N_{v_{\parallel}},N_{\mu})=(32, 24, 16, 8)$; the orange curves are for $(N_x, N_z, N_{v_{\parallel}},N_{\mu})=(32, 96, 32, 16)$.}
\end{figure*}

We have run the simulations for $t=2.0~\mathrm{ms}$ until they reach an equilibrium. Importantly, we found the simulation with the explicit LBD is not well converged at the original phase space resolution $(N_x, N_z, N_{v_{\parallel}},N_{\mu})=(32, 24, 16, 8)$. Utilizing a higher resolution in $z$ and in velocity space, i.e., $(N_x, N_z, N_{v_{\parallel}},N_{\mu})=(32, 96, 32, 16)$, we were able to obtain better convergence with the explicit LBD simulation, getting better agreement with the implicit BGK results. Similarly, we have run implicit BGK simulations at the low and the high resolutions for comparison. Figure~\ref{fig: profiles} is a panel of equilibrium profiles for four simulations using the explicit LBD and implicit BGK at two different resolutions. The three columns correspond to three types of profiles: parallel profiles at the radial center of the domain $x=x_c=0.162$, radial profiles at the inboard midplane $z=\pi/2$, and radial profiles at the outboard midplane $z=-\pi/2$. The four rows correspond to four physical quantities: electron density, electron temperature, ion temperature, and electric potential. For the explicit LBD, the parallel and radial profiles of electron temperature and electric field change significantly as the resolution increases. In particular, the electron temperature oscillates in the parallel direction, and it is higher on the low field side at the low resolution as shown in Fig.~\ref{fig: profiles}(b). For the implicit BGK, by contrast, all these quantities exhibit profiles that are very close at both resolutions, indicating that the implicit BGK converges at low resolution. Moreover, the equilibrium profiles of all these quantities are in good agreement for the two different collision operators at the high resolution, thereby validating our implicit BGK collision operator. 

The time step in the explicit LBD runs is limited by the fastest time scale in the collision operator, $\Delta t_{\rm coll} \sim (\Delta v_\parallel/v_t)^2 / \nu$. (The diffusion in the $\mu$ direction sets a comparable time scale.) The implicit BGK operator is no longer limited by the collision rate, but there are two other rates that still need to be considered. First, the code estimates a Courant limit on the time step due to the high-frequency electrostatic shear Alfvén mode $\Omega_{\rm h}$ mode\cite{Lee_1983}, $\Delta t\Omega_{\rm h, max}<CFL$. This mode is essentially the $\beta \rightarrow 0$ electrostatic limit of an Alfven wave that exists in gyrokinetic theory, where it has the dispersion relation $\Omega_{\rm h} = k_\parallel v_{te}  /k_\perp \rho_s$ in a simple limit with $\omega \gg k_\parallel v_{te}$. The maximum frequency of this mode is given by $\Omega_{\rm h, max}\approx k_{\parallel \rm max} v_{te} / k_{\perp \rm min} \rho_s$. Second, the time step limit due to advection on the grid, which is typically set by the fastest electrons along the field lines, so $\Delta t < \min(\Delta z) / v_{e, \rm max}$. For this specific ASDEX case, the $\Omega_{\rm h}$ mode is the faster rate, and sets a time step $\Delta t$ that is only about 2 times larger than the collisional time step for the explicit LBD case. At present the implicit BGK runs are about twice as expensive per time step as the LBD runs, so the net result is that implicit BGK and LBD runs at the same resolution have about the same wall-clock run time for these ASDEX parameters. The expected speedup at fixed resolution scales as $
\nu R  / v_t \sim n_e R / T_e^2$, and so would give larger speed-ups at higher collisionality, such as in reactor designs near divertor detachment.

There are ways to improve this significantly.  There is a method to replace the current iterative method of constructing the Maxwellians in the BGK operator in a conservative way that is the equivalent of 3 iterations instead of the present $10$ iterations.  This would speed up the code by about a factor of 2 (including the cost of the Hamiltonian terms).  There are also semi-implicit methods that should be able to step over the fast $\Omega_{\rm h}$ mode frequency, so that the parallel electron motion becomes the time limit, which is a factor of 9.6 times larger.  Together, these additional algorithmic improvements could make an implicit BGK run about 10-20 times faster than an explicit LBD run at the same resolution.

Even without further algorithm improvements, there is a big advantage of using the implicit BGK operator instead of the LBD operator for this case because it is more robust on a coarse grid and does not need as much resolution as the LBD operator to converge. At least in part this is probably because the BGK operator does not involve taking derivatives. Going from the high resolution grid of $(N_x, N_z, N_{v_\parallel}, N_\mu) = (32, 96, 32, 16)$ to a moderate resolution grid of $(N_x, N_z, N_{v_\parallel}, N_\mu) = (32, 24, 16, 8)$ (what we call \enquote{low resolution} elsewhere in this paper) reduces the number of grid points by a factor of 16. There is another factor of $\sim 4$ increase in the time step as $\Delta z$ increases (this is not exact because the grid is non-uniformly spaced in the physical parallel coordinate $z$), giving a total ideal speedup of about 64. The actual speed-up observed in going from high resolution LBD to moderate resolution BGK in this case is a factor of 56. (It might be possible to improve the LBD operator by using flux limiters or positivity-preserving interpolation methods to make it more robust and allow it to work better on coarser grids.) Note that the resolution numbers here are the number of DG cells in each direction. For approximate comparison with grid codes, they should be doubled in all directions, except tripled in $N_{v_\parallel}$, because $p=2$ basis functions are used in that direction.

Next, we can investigate the parallel heat flux 
\begin{equation}
    q_{\parallel s} = \frac{1}{2}m_s\int v^2v_{\parallel}f_s d\mathbf{v}
\end{equation}
of electrons and ions for these simulations. Figure.~\ref{fig: heat flux profiles} shows the electron and ion heat flux profiles on the inner and outer divertor plates for the four comparative cases. Both the ion and electron heat flux are larger on the outer divertor plate than the inner plate. As in Fig.~\ref{fig: profiles}, we find that the divertor heat fluxes have converged fairly well at low resolution for the implicit BGK case, but the LBD case needs higher resolution. The vertical scales for the electron heat flux plots were chosen to see the values better for $\psi \gtrsim 0.155$. Near the inner $\psi$ boundary, the electron heat flux gets artificially very large because we use a $\phi=0$ boundary condition there, which turns off the sheath potential there. We will consider improved inner boundary conditions for $\phi$ in future work. These initial SOL simulations are aimed at studying the far SOL, away from the inner boundary.

\section{\label{sec: conlusions}Conclusions and discussion}
We have implemented an implicit BGK collision operator in the framework of Gkeyll to speed up simulations in the highly collisional SOL. We employ a Picard fixed-point iteration (quasi-Newton with identity Jacobian) to enforce moments conservation of the discretized Maxwellian that appears in the BGK operator. The iteration efficiency is demonstrated by a benchmark test that the first three velocity moments of the Maxwellian converge near machine precision within $\cO(10)$ iterations. In this implementation, we utilize a first order split which combines a standard Strong Stability Preserving (SSP) third-order Runge-Kutta (RK3) explicit method for the collisionless advection with an unconditionally stable backward Euler method for the BGK collision operator. [There has been interesting work on higher-order splitting or IMEX methods in recent years, exploring methods that are not only higher-order accurate but also positivity preserving or asymptotic preserving to various orders of accuracy, see for example \cite{HuZhang_2017, DegondDeluzet_2017, Gottlieb_2022} and references therein. These could be explored in the future.] This removes the constraint on the time step from the collision term. Utilizing this implementation leads to significant speed-ups in the high collisionality limit: the Sod shock test runs 7480 times faster without losing the characteristics in the results. This implicit BGK collision operator has also been generalized to multispecies collisions. This feature is validated by comparison with the explicit LBD operator in the relaxation test of an anisotropic deuterium plasma. Overall, the two collision operators are in good agreement on equilibration of the mean parallel velocity. For the relaxation of temperature, we observe minor differences in the isotropization of $T_{\parallel}$ and $T_{\perp}$ as expected first, but then the evolution becomes identical on the interspecies energy-exchange timescale. 

As a demonstration of the utility of this approach, we have applied the implicit BGK collision operator to axisymmetric simulations of the scrape-off layer (SOL) in an ASDEX Upgrade (AUG) plasma, solving the gyrokinetic equation in a field line following coordinate system generated from a numerical equilibrium. In the transport steady state, the profiles of electron density, electron temperature, ion temperature and electric potential agree well with those from simulations with the explicit LBD operator at high resolution, $(N_x, N_z, N_{v_{\parallel}},N_{\mu})=(32, 96, 32, 16)$. While the speed of the BGK and LBD runs are comparable for this ASDEX case at the same resolution, the BGK operator is more robust and converges at lower resolution, giving a net speedup of a factor of 56 at $(N_x, N_z, N_{v_{\parallel}},N_{\mu})=(32, 24, 16, 8)$. (These numbers are the number of DG cells in each direction. An equivalent grid is doubled in all directions except tripled in $N_{v_\parallel}$.) The implicit BGK speedup might be improved by another factor of 10-20 with additional algorithmic improvements discussed in Sec.~\ref{sec: axisymmetric sims}.

As discussed in the introduction, in future work the BGK operator could be made significantly more accurate for high collisionality regimes by making the collision frequency vary with velocity, to have the key $\nu \sim 1/v^3$ behavior at high velocity.  This will give more accurate values for the transport coefficients, as shown for gases in \cite{Struchtrup_1997, Mieussens_2004, Haack_2021, Haack_2023}. One could further improve this with a hybrid method, where an explicit LBD or full Rosenbluth Fokker-Planck operator is used in most of the plasma, where the collision frequency is low, while transitioning to an implicit BGK operator in regions of higher collision frequency, such as in the SOL or near divertor plates.

In this work, the implicit BGK has already achieved significant speed-up with 2D axisymmetric simulations of AUG SOL. Given the success of this implicit BGK implementation and its favorable comparisons to more accurate collision operators in this higher collisionality limit, future work will apply the implicit BGK operator to 3D nonlinear turbulence simulations of AUG SOL to study the turbulent transport of blob filaments. AUG experiments in L-mode have reported a correlation between the transition of filamentary regimes and the flattening of the SOL density profiles referred to as \enquote{shoulder formation}\cite{Carralero_2014}. Results of the experiments were consistent with theoretical models describing the process as a transition between the conduction-dominated sheath-limited (SL) regime and the convection-dominated inertial (IN) regime triggered by increased parallel collisionality \cite{Carralero_2015}. Later experiments identified that the trigger was specifically the increased collisionality at the divertor \cite{Carralero_2017}. Further AUG experiments in L-mode reported an increase of the radial convective particle flux associated to filaments by almost an order of magnitude after the shoulder formation \cite{Carralero_2018}. By extending our simulations with realistic AUG geometry to 3D, we will closely investigate the turbulence transport before and after the blob regime transition. And we emphasize that the results of this paper show a path forward for simulating plasmas in this regime, avoiding the stiff collisional transport time scales which would preclude explicit time integration that not only requires a smaller time step, but as has been shown here, also requires higher resolution to properly resolve the dynamics.

\begin{acknowledgments}
We thank Grant Johnson for his assistance with the implementation of the implicit time integrator for the BGK operator. This work is supported by a DOE Distinguished Scientist award, the CEDA SciDAC project and other PPPL projects via DOE Contract Number DE-AC02-09CH11466 for the Princeton Plasma Physics Laboratory. This research used resources of the National Energy Research Scientific Computing Center (NERSC), a Department of Energy User Facility using NERSC project allocation 4907.
\end{acknowledgments}


%
%

%


\bibliography{refs}

\end{document}